\documentclass[pra,twocolumn,showpacs,floatfix,a4paper,superscriptaddress]{revtex4}
\usepackage{bm,color,graphicx,amsmath,txfonts}
\usepackage{hyperref}

\newcommand{\abs}[1]{\left|{#1}\right|}

\newcommand{\av}[1]{\left\langle #1 \right\rangle}

\newcommand{\pt}[1]{\left( #1 \right)}

\newcommand{\ii}{{\rm i}}

\newcommand{\FF}{{\cal F}}

\newcommand{\NN}{{\cal N}}

\begin{document}

\title{Probing Spontaneous Wave-Function Collapse with Entangled Levitating Nanospheres}

\author{Jing Zhang}
\affiliation{State Key Laboratory of Quantum Optics and Quantum Optics Devices, Institute of Opto-Electronics, Shanxi University, Taiyuan 030006, China}
\affiliation{Collaborative Innovation Center of Extreme Optics, Shanxi University, Taiyuan 030006, China}
\author{Tiancai Zhang}
\affiliation{State Key Laboratory of Quantum Optics and Quantum Optics Devices, Institute of Opto-Electronics, Shanxi University, Taiyuan 030006, China}
\affiliation{Collaborative Innovation Center of Extreme Optics, Shanxi University, Taiyuan 030006, China}
\author{Jie Li}
\affiliation{State Key Laboratory of Quantum Optics and Quantum Optics Devices, Institute of Opto-Electronics, Shanxi University, Taiyuan 030006, China}
\affiliation{Collaborative Innovation Center of Extreme Optics, Shanxi University, Taiyuan 030006, China}
\affiliation{School of Science and Technology, Physics Division, University of Camerino, I-62032 Camerino (MC), Italy}

\begin{abstract}
Wave function collapse models are considered as the modified theories of standard quantum mechanics at the macroscopic level. By introducing nonlinear stochastic terms in the Schr\"odinger equation, these models make predictions, differently from those of standard quantum mechanics, that it is fundamentally impossible to prepare macroscopic systems in macroscopic superpositions. The validity of these models can only be examined by experiments and hence efficient protocols for this kind of experiments are highly needed. Here we provide a protocol that is able to probe the postulated collapse effect by means of the entanglement of the center-of-mass motion of two nanospheres optically trapped in a Fabry-P\'erot cavity. We show that the collapse noise results in large reduction of the steady-state entanglement and the entanglement, with and without the collapse effect, shows distinguishable scalings with certain system parameters, which can be used to unambiguously determine the effect of these models.

\end{abstract}


\date{\today}
\maketitle

\section{INTRODUCTION}

Wave function collapse models (CMs)~\cite{BassiRMP}, as the modified theories of standard quantum mechanics, postulate a fundamental breakdown of quantum superposition at the macroscopic scale. They have been proposed to explain the lack of observations of macroscopically distinguishable superposition states of macroscopic objects and offer a possible solution to the quantum measurement problem~\cite{Bassi} and the quantum-to-classical transition. There are several different versions of the CMs, e.g., the Ghirardi-Rimini-Weber approach~\cite{GRW}, continuous spontaneous localization (CSL)~\cite{CSL}, and gravitationally-induced CMs~\cite{DP}. These models modify the Schr\"odinger equation by introducing appropriate stochastic nonlinear terms, of which the effect is negligible for microscopic systems, while becomes prominent for macroscopic objects resulting in the emergence of macroscopic classicality.

Many proposals have been put forward to examine these CMs in different systems and with different approaches. In general, they can be divided into two kinds: in an interferometric way and in a noninterferometric way. For the former, matter-wave interferometry~\cite{ArndtRMP} is typically used, where large massive objects, e.g., molecules or clusters~\cite{cluster}, are sent through interference gratings. However, to date the mass range has not been reached to effectively test the CMs. For the latter, protocols~\cite{Mauro,Nimmrichter,Sekatski,Diosi,Jie,Barker0} have been recently provided based on cavity optomechanics~\cite{OMRMP}. The main advantage of this approach is that the preparation of large spatial superposition states~\cite{mirror,nanosphere,Ulbricht,Wan} is not required. It has been shown~\cite{Mauro,Nimmrichter,Sekatski,Jie} that collapse noise induced momentum diffusion of a mechanical resonator (MR) could be probed in the phase noise of the cavity output light. Apart from the above, more recently quantum estimation theory has been applied to discriminate the effect of CMs~\cite{Serafini,Millen}. 

In this paper, we provide a novel scheme to test the CMs by means of the steady-state entanglement of two macroscopic MRs. Many protocols have been proposed for the generation of entanglement between two massive MRs using optomechanical and/or electromechanical systems, e.g., by exploiting radiation pressure~\cite{PRL02,jopa,genesNJP,hartmann}, by transferring entanglement~\cite{Peng03,Jie13,Ge13} or squeezing~\cite{EPL} from optical fields, by conditional measurements on light modes~\cite{entswap,bjorke,mehdi1,woolley,mehdi2,Savona}, and by reservoir engineering realized by properly choosing multi-frequency drivings~\cite{Clerk,Tan,WoolleyClerk,Abdi2,Buchmann,JieNJP,JieCF}. We will focus on the CSL model which is one of the most widely studied CMs. The scheme is based on the known fact that entanglement, as a kind of quantum correlations, is particularly sensitive to various noises. A small rise of noise may significantly degrade the entanglement. As is known, the collapse noise postulated in the CMs is typically very small thus bringing the challenges for experimental verification. In view of these, entanglement could act as a perfect probe that may be able to sense whether the collapse noise is present or not. The reason we adopt {\it steady-state} entanglement is that, in order to efficiently test the CMs, one should prepare entangled states lasting for a relatively long time~\cite{Belli,Singh}. We know that entanglement in steady states, rather than in transient states, is harder to prepare due to the continuous decoherence process interacting with various noises.  The effect of the CMs would be more noticeable in the steady-state entanglement due to the time accumulation effect.

The system used to test the collapse theories should possess as small as possible environmental noises, comparable to the hypothetical collapse noise, and the test object should be large or massive enough yielding considerable collapse effect. Levitated nanospheres~\cite{nanosphere,Chang,Barker,Raizen,gieseler,Markus}, owning very high mechanical quality factors, could be the ideal platform to implement such a test. Protocols~\cite{Jie,Barker0} have been provided using a single nanosphere trapped in a Fabry-P\'erot optical cavity that are able to test the strength of the collapse rate in the CSL theory to values as low as $10^{-12}$ Hz with realistic parameters. In the present paper, here we study the effect of the CSL on quantum correlations of two macroscopic objects. Specifically, we employ two nanospheres optically trapped in a Fabry-P\'erot cavity and study the CSL effect on the stationary entanglement of the center-of-mass motion of the two spheres. Since the entanglement is particularly sensitive to noises, the diffusion rates of various noises in the system must be small. We find that, with properly chosen parameters, the collapse noise results in large reduction of the entanglement and the entanglement shows distinguishable scalings, with and without the CSL effect, with certain system parameters, e.g., the trapping frequency, due to the fact that different sources of noise exhibit different scalings with the system parameters~\cite{Jie}. The above observation can unambiguously determine whether the collapse noise is or is not actually present or effective.

The remainder of the paper is organized as follows. In Sec. \ref{system} we describe in detail our system, provide the quantum Langevin equations for achieving the entanglement of the two nanospheres, and analyze the diffusion rates of various noises (both environmental and the postulated collapse noise) in the system. In Sec. \ref{results} we present the results and show the details of the parameters with which the CSL effect with the corresponding value of the collapse rate could be determined. Finally, Sec. \ref{concl} is devoted to conclusions and discussions and the Appendix gives the details for calculating the steady-state entanglement.

\begin{figure}[t]
\centering
\includegraphics[width=3.3in]{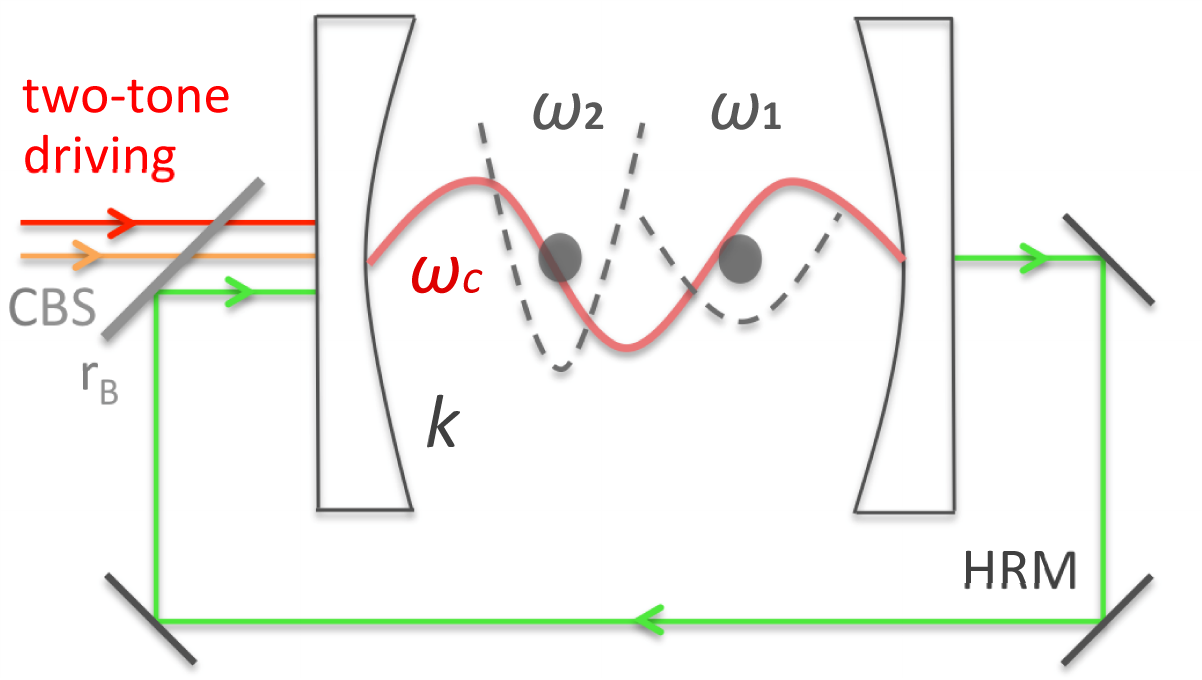}
\caption{Sketch of the proposed scheme for testing the CSL theory with entangled nanospheres. Two identical nanospheres are optically trapped with different frequencies $\omega_1$ and $\omega_2$ in a Fabry-P\'erot cavity. The cavity mode is bichromatically driven at the two frequencies $\omega_0+\omega_1$ and $\omega_0-\omega_2$. Large and robust entanglement between the two spheres can be generated in the steady state. The output field of the cavity is fed back into the input port through highly reflective mirrors (HRM) and a controllable beamsplitter (CBS) with tunable reflection coefficient $r_B$, which is used for reducing the cavity loss and improving the entanglement.}
\label{fig1}
\end{figure}

\section{The system}
\label{system}

Before making the test, we shall first prepare two nanospheres into entangled states. We adopt the scheme provided in Ref.~\cite{JieCF} which is able to generate large and robust entanglement between two MRs in steady states. As depicted in Fig.~\ref{fig1}, we consider two identical spheres of radius $R$ trapped by harmonic dipole traps with different frequencies $\omega_1$ and $\omega_2$ in two different potential wells within a Fabry-P\'erot cavity. The center-of-mass motion of the sphere is modeled as a quantum-mechanical harmonic oscillator. The cavity is of length $L$, finesse $\FF$, and mirror radius of curvature $R_c$. A single cavity mode with resonance frequency $\omega_c$, interacting via the usual optomechanical interaction with the two MRs, is bichromatically driven, with powers $P_1$ and $P_2$, at the two sideband frequencies $\omega_{L1} =\omega_0 +\omega_1$ and $\omega_{L2}=\omega_0-\omega_2$ with the reference frequency $\omega_0$ detuned from the cavity resonance by $\Delta_0=\omega_c-\omega_0$. This means that the cavity mode is simultaneously driven at the blue sideband associated with the 1st MR and at the red sideband associated with the 2nd MR. We note that the scheme~\cite{JieCF} is the improved version of the one~\cite{JieNJP} by introducing a coherent feedback loop, which leads to a significantly reduced effective cavity decay rate and a remarkable improvement of the entanglement. This is vital and makes it possible to test the CMs based on this scheme because the mechanical frequency $\omega_{1,2}$ now can take much smaller values (since the effective cavity decay rate $\kappa_{\rm eff}$ is significantly reduced due to the feedback) to fulfill the condition of the scheme $\kappa_{\rm eff} \ll \omega_{1,2}, \, |\omega_1 -\omega_2|$~\cite{JieNJP}. As shown in Ref.~\cite{Jie}, the diffusion rate $D_t$ due to the scattering of trapping light, which will be the main diffusion for the system at low pressure and temperature, is proportional to the trapping frequency $\omega_{1,2}$. A much smaller $\omega_{1,2}$ yields a much smaller diffusion rate $D_t$, making it possible to generate sizable stationary entanglement between the two MRs.

\begin{figure*}[t]
\hskip1.1cm{\bf (a)}\hskip5.28cm{\bf (b)}\hskip5.25cm{\bf (c)} \\
\includegraphics[width=0.308\linewidth]{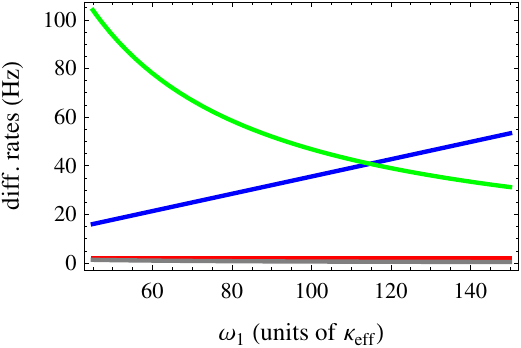}
\hskip0.1cm\includegraphics[width=0.308\linewidth]{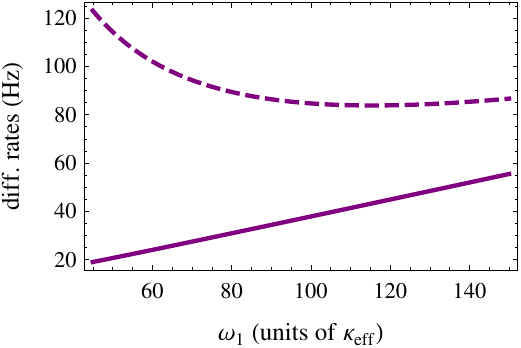}
\hskip0.1cm\includegraphics[width=0.302\linewidth]{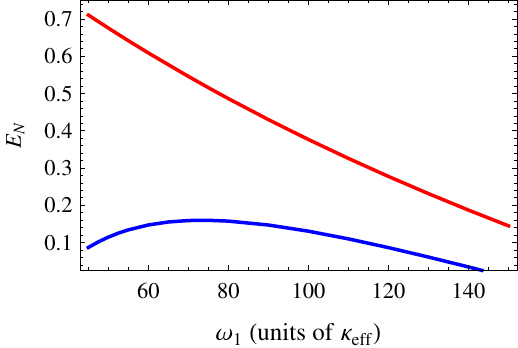}   \\
\caption{(a) Diffusion rates for the scattering of trapping light $D_t^1$ (blue), cavity light $D_c^1$ (red), air molecules $D_a^1$ (gray), and the collapse rate $\lambda_{\rm sph}^1$ (green) versus the trapping frequency $\omega_1$. Note that the curves of $D_c^1$ and $D_a^1$ are very close to the $\omega_1$-axis and no longer visible. (b) The solid line denotes the sum of the diffusion rates $D_t^1$, $D_c^1$ and $D_a^1$ in (a), while the dashed line includes also the collapse rate $\lambda_{\rm sph}^1$ on the basis of the solid line. (c) Steady state entanglement $E_N$ between the two nanospheres versus the trapping frequency $\omega_1$. Blue (red) line refers to the case with (without) the CSL effect. The parameters are $R=0.15r_c$, $r_B=0.99$, $G_2=1.2\kappa_{\rm eff}$, $G_1=0.72G_2$, $\omega_2=2\omega_1$, $L=4$ cm, $\kappa=20$ kHz (corresponding to $\FF=5.9\times10^5$ and $\kappa_{\rm eff}=200$ Hz), $R_c=L/1.5$, $\lambda_c=1064$ nm, $\NN=0.8$, $T=10$ mK, $P_a=10^{-12}$ Torr, $\lambda=10^{-8}$ Hz, $r_c=100$ nm, and we use diamond nanospheres with $\rho_0=3.5$ g/cm$^3$ and $\epsilon=5.76$.}
\label{fig2}
\end{figure*}

The system dynamics can be efficiently studied by linearizing the optomechanical interaction in the limit of large driving field. The relevant degrees of freedom for the linearized dynamics are the fluctuations of the cavity field and of the mechanical center-of-mass variables about their respective average values, described by the amplitude and phase quadratures $X$ and $Y$ (with $[X,Y]{=}\ii$) of the cavity field, and by the dimensionless position and momentum $x_j$ and $p_j$ (with $[x_j,p_j]{=}\ii$, $j{=}1,2$) of the nanospheres. The corresponding quantum Langevin equations, in the interaction picture with respect to $H_0{=}\hbar\omega_0 (X^2{+}P^2)/2+\hbar\sum_{j=1,2}\omega_j (x_j^2{+}p_j^2)/2$, are given by~\cite{JieNJP,JieCF}
\begin{equation}
\begin{split}
\dot X&=-\kappa_{\rm eff}\, X-G_1\,p_1+G_2\,p_2+\sqrt{2\kappa_{\rm eff}}\,X^{\rm in},\\
\dot Y&=-\kappa_{\rm eff}\, Y-G_1\,x_1-G_2\,x_2+\sqrt{2\kappa_{\rm eff}}\,Y^{\rm in},\\
\dot{x}_j&=-\frac{\gamma}{2} x_j+(-1)^j G_j\, Y+F_{x}^j,\\
\dot{p}_j&=-\frac{\gamma}{2} p_j-G_j\, X+F_{p}^j,\\
\end{split}
\label{QLEs}
\end{equation}
where we have set the effective detuning $\Delta\,{=}\,\Delta_0\,{+}\,\delta$ ($\delta$ the frequency shift due to the optomechanical interaction and also the feedback) equal to zero, which is the optimal detuning for the generation of entanglement~\cite{JieCF}. $\kappa_{\rm eff}\,{=}\,\kappa\,(1{-}|r_B| \cos \theta)$ is the effective cavity decay rate due to the coherent feedback where $r_B$ is the reflection coefficient of the controllable beamsplitter in the feedback loop (see Fig.~\ref{fig1}), $\theta$ is the phase shift of the light in the feedback loop, and $\kappa=\pi c/(2 \FF L)$ ($c$ the speed of light) is the cavity decay rate without feedback. We see that $\kappa_{\rm eff}$ can be significantly reduced when $|r_B| \to 1$ and $\theta=2n\pi$ ($n=0,1,2,...$). In practice, one can always adjust this phase shift and set it equal to the optimal value for the entanglement $\theta=2n\pi$~\cite{JieCF}. $G_j{=}g_j \alpha_j$ is the effective optomechanical coupling where $\alpha_j=\sqrt{2\kappa_{\rm eff} P_j/[\hbar\omega_{Lj}(\omega_j^2+\kappa_{\rm eff}^2)]}$ and $g_j{=}\omega_c\sqrt{\frac{\hbar}{m\omega_j}}\frac{2\pi}{\lambda_c}\frac{\epsilon-1}{\epsilon+2}\frac{3V_s}{4V_c}$~\cite{Chang} is the bare optomechanical coupling associated with the $j$-th MR, with $\lambda_c$ the cavity wavelength, $\epsilon$ the electric permittivity of the sphere, $V_s$ its volume, and $V_c{=}\pi L W_0^2/4$ the cavity mode volume with mode waist $W_0=\sqrt{\lambda_c L (2R_c/L{-}1)^{1/2}/2\pi}$. $\gamma=\frac{16}{\pi}\frac{P_a}{\bar v\,R\,\rho_0}$ is the mechanical damping rate due to the friction with residual air molecules, with $P_a$ the gas pressure, $\rho_0$ the mass density of the sphere, $\bar v=\sqrt{3k_B T/m_a}$ the mean speed of the air molecules, $m_a$ their mass (which we take $m_a=28.97$ amu), and $T$ the gas temperature~\cite{Chang}. For levitated nanospheres $\gamma$ can be very small leading to very high quality factors, $\gtrsim10^{10}$~\cite{Oriol2011}. $X^{\rm in}$ and $Y^{\rm in}$ are the quadratures of the vacuum noise entering into the cavity and their only nonzero correlation functions are 
\begin{equation}
\av{X^{\rm in}(t)\,X^{\rm in}(t')}=\av{Y^{\rm in}(t)\,Y^{\rm in}(t')}=\frac{1}{2} \delta(t-t'). 
\end{equation}
$F_x^j$ and $F_p^j$ are the combined force operators in the rotating frame which include all the relevant stochastic forces accounting for the mechanical diffusion. The only nonzero correlation functions are
\begin{equation}
\begin{split}
\av{F_x^j(t)\,F_x^{j'}(t')}&=\av{F_p^j(t)\,F_p^{j'}(t')}  \\
&=\frac{1}{2} \delta_{j,j'}\,\delta(t{-}t')\pt{D_a^j+D_t^j+D_c^j+\lambda_{\rm sph}^j},   \\
\end{split}
\end{equation}
where $D_a^j$, $D_t^j$, $D_c^j$ and $\lambda_{\rm sph}^j$ are, respectively, the diffusion rates caused by the scattering of background air molecules, of trapping light, of cavity photons, and by the collapse noise. In the relevant high temperature limit, $D_a^j$ is given by
\begin{equation}
D_a^j=2\gamma\, \frac{k_B T}{\hbar\omega_j}.
\end{equation}

The diffusion rates due to the scattering of trapping and cavity light are given, respectively, by~\cite{Pflanzer}
\begin{equation}
D_t^j=\frac{8\epsilon_c^2 k_c^6 R^3}{9 \rho_0 \omega_j}\frac{{\cal I}_j}{\omega_{Lt}}\ ,
\hspace{1cm}
D_c^j=\frac{2\epsilon_c^2 k_c^6 R^3}{9 \rho_0 \omega_j}\frac{\hbar n_{\rm ph}c}{V_c}\ ,
\label{diff}
\end{equation}
where $\epsilon_c{=}3\frac{\epsilon-1}{\epsilon+2}$, $k_c=2\pi/\lambda_c$, $\omega_{Lt}$ is the frequency of the trapping laser, ${\cal I}_j$ is the intensity of the trapping field, which is given by ${\cal I}_j{=}P_{tj}/(\pi W_t^{2})$ with $P_{tj}$ the laser power and $W_t$ the waist of trapping light which is approximated by $W_t\approx\lambda_c/(\pi{\cal N})$ with ${\cal N}$ the numerical aperture, and the trapping frequency is determined by $\omega_j=[4\epsilon_c {\cal I}_j/(\rho_0 c W_t^{2})]^{1/2}$. Finally, $n_{\rm ph}=\abs{\alpha_1}^2+\abs{\alpha_2}^2$ is the mean cavity photon number.

According to the CSL theory, the collapse noise induced diffusion rate for a spherical particle is given by~\cite{Nimmrichter}
\begin{equation}
\lambda_{\rm sph}^j=\frac{\hbar}{\omega_j}\frac{8\pi\,\lambda\,\rho_0}{m_0^2}\left[e^{-R^2/r_c^2}-1+\frac{R^2}{2r_c^2}(e^{-R^2/r_c^2}+1)\right]\frac{r_c^4}{R^3} \\ ,
\label{lambda_sph}
\end{equation}
with $m_0$ the atomic mass unit.
The actual strength of collapse noise is determined by two phenomenological parameters:
the characteristic length $r_c$ and the collapse rate $\lambda$. The characteristic length is typically set at $r_c\simeq 100$ nm, above which the collapse effect tends to be prominent. $\lambda$ denotes the average collapse rate at one proton mass. The initial estimate of $\lambda$ is $10^{-16}$ Hz~\cite{GRW,CSL}, while larger values have been proposed, e.g., $10^{-8\pm2}$ Hz given by Adler~\cite{Adler}. Up to now, different experiments have indicated that $\lambda$ should be lower than ${\sim}10^{-8}$ Hz~\cite{Vinante,coldatoms,Chen,Vinante2,VinanteNEW}, ${\sim}10^{-9}$ Hz~\cite{Pearle2014} and ${\sim}10^{-11}$ Hz~\cite{Curceanu}, for $r_c\simeq 100$ nm. Does it exist an exact value or range of $\lambda$ and, if exists, how large is it? These questions can only be answered by experiments.

\begin{figure*}[t]
\hskip0.8cm{\bf (a)}\hskip5.2cm{\bf (b)}\hskip5.2cm{\bf (c)} \\
\hskip0.1cm\includegraphics[width=0.287\linewidth]{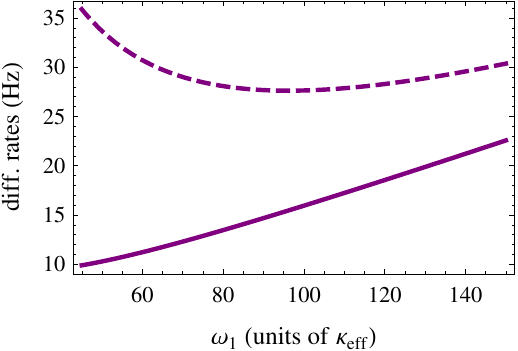}
\hskip0.34cm\includegraphics[width=0.287\linewidth]{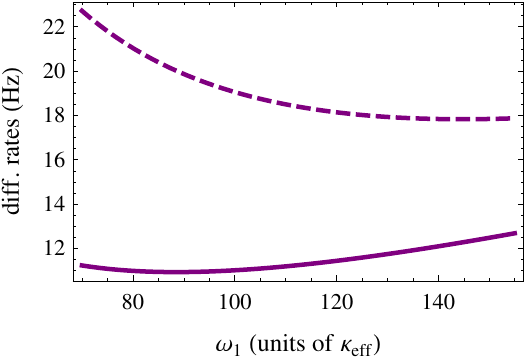}
\hskip0.315cm\includegraphics[width=0.298\linewidth]{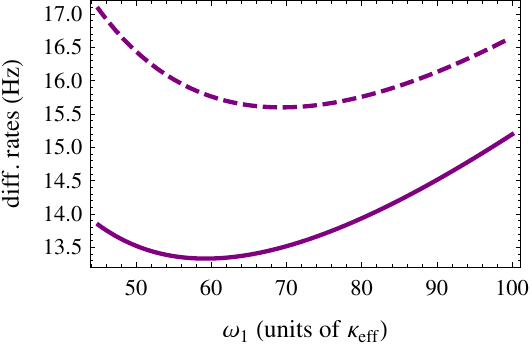}   \\
\hskip0.8cm{\bf (d)}\hskip5.2cm{\bf (e)}\hskip5.2cm{\bf (f)} \\
\includegraphics[width=0.3\linewidth]{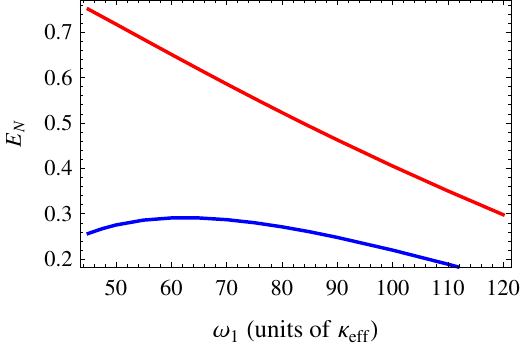}
\hskip0.1cm\includegraphics[width=0.3\linewidth]{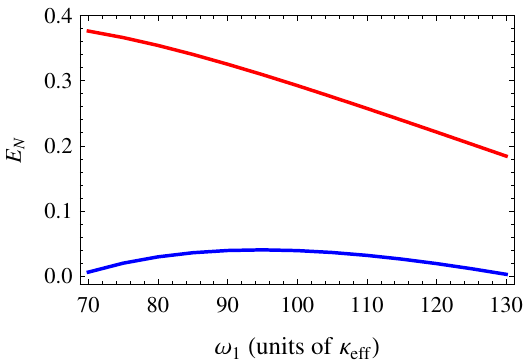}
\hskip0.1cm\includegraphics[width=0.3\linewidth]{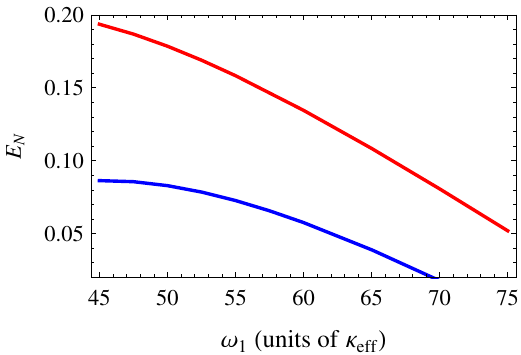}  \\
\caption{(a)-(c) Total diffusion rates with (dashed line, i.e., $D_t^1+D_c^1+D_a^1+\lambda_{\rm sph}^1$) and without (solid line, i.e., $D_t^1+D_c^1+D_a^1$) the CSL effect versus the trapping frequency $\omega_1$. (d)-(f) Steady state entanglement $E_N$ of the two nanospheres versus the trapping frequency $\omega_1$. Blue (red) line corresponds to the case with (without) the CSL effect. The parameters are as follows: (a), (d) $\lambda=10^{-9}$ Hz, $R=0.15r_c$, $r_B=0.996$, $G_2=1.2\kappa_{\rm eff}$, $G_1=0.77G_2$. (b), (e) $\lambda=10^{-10}$ Hz, $R=0.18r_c$, $r_B=0.999$, $G_2=2.2\kappa_{\rm eff}$, $G_1=0.79G_2$. (c), (f) $\lambda=10^{-11}$ Hz, $R=0.22r_c$, $r_B=0.999$, $G_2=2\kappa_{\rm eff}$, $G_1=0.79G_2$. The other parameters are as in Fig.~\ref{fig2}.}
\label{fig3}
\end{figure*}

\section{Test of the CSL theory with entangled nanospheres}
\label{results}

\begin{table}[b]
\caption{Different scalings of the diffusion rates $D_t$, $D_c$, $D_a$ and $\lambda_{\rm sph}$ with two key parameters of the system, i.e., the radius of the sphere $R$ and the trapping frequency $\omega$. Symbol $``\uparrow"$ ($``="$) denotes an increasing (constant) function of the parameter. }
\label{table}
\begin{tabular}{ p{1.4cm} | p{1.1cm} |  p{1.88cm} | p{1.2cm} | p{2.3cm} }    \hline 
Parameter      & $\,\,\,\,\,\,\,D_t$   & $\,\,\,\,\,\,\,\,\,\,\,\,\,\,D_c$   & $\,\,\,\,\,\,\,\,D_a$   & $\,\,\,\,\,\,\lambda_{\rm sph}$ \\	 \hline 
 $\,\,\,\,\,\,\,\,\,R$   & $\,\,\,\,\,\propto R^3$   & $\,\,\,\,=\,$ (fixed $G$)  & $\,\,\,\,\,\propto R^{-1}$    & $\,\,\,\,\,\,\,\uparrow\,$ ($R<2.38 r_c$)   \\    \hline
$\,\,\,\,\,\,\,\,\,\omega$   & $\,\,\,\,\,\propto \omega$   &$\,\,\,\,=\,$ (fixed $G$)   & $\,\,\,\,\,\propto \omega^{-1}$  & $\,\,\,\,\,\,\propto \omega^{-1}$   \\     \hline
\end{tabular}
\end{table}

We observe that $D_t$, $D_c \propto R^3$ ($D_c$ is, however, independent with $R$ for fixed values of $G$), $\lambda_{\rm sph}$ increases with $R$ when $R<2.38 r_c$, while $D_a \propto 1/R$, as displayed in Table~\ref{table}. Since the entanglement is particularly sensitive to various noises, implying that a relatively small size of the sphere should be adopted, we consider the radius of the sphere $R=0.15$--$0.22$ $r_c$ (while a larger size $R=r_c$ has been used in Ref.~\cite{Jie} where the observable is instead the phase quadrature of the output light), which yields not large $D_t$, $D_c$, $D_a$ and meantime comparable $\lambda_{\rm sph}$, as shown in Fig.~\ref{fig2} (a). We have taken as small as possible the value of $G_{2}$ (the ratio of $G_1/G_2$ is however optimized for the entanglement~\cite{JieCF,JieNJP}), which gives a negligible $D_c$ but is large enough to generate sizable entanglement. Another reason $G_{1,2}$ must be small is that in order to make the scheme of Ref.~\cite{JieCF} valid $G_{1,2} \ll \omega_{1,2}, |\omega_1-\omega_2|$ must be fulfilled~\cite{JieNJP} . We have also assumed the system is at low temperature $T=10$ mK and pressure $P_a=10^{-12}$ Torr, resulting in a sufficiently small $D_a$. Figure~\ref{fig2} is plotted as a function of the mechanical frequency $\omega_{1,2}$ which is a key parameter~\cite{Jie} and can be easily altered by adjusting the intensity of the trapping laser. Under these conditions, one could see two distinguishable curves of the total diffusion rate with and without the CSL effect (see Fig.~\ref{fig2} (b)) which thereby result in distinguishable curves of the entanglement (see the Appendix for calculating the steady-state entanglement), as shown in Fig.~\ref{fig2} (c). We see that the entanglement in the absence of the CSL effect decreases almost linearly with $\omega_{1,2}$, while it increases first and then decreases with the CSL effect, and more noticeably, the difference of their values in these two cases is quite large confirming the sensitivity of the entanglement to the noise. Therefore, the CSL effect could be determined by repeating the experiment at different trapping laser powers and verifying the distinguishable behavior, particularly the different signs of the slope of the curves, in the region of small $\omega_{1,2}$.

In Fig.~\ref{fig3} we present the results for testing the CSL with different collapse rates $\lambda=10^{-9}$ Hz, $10^{-10}$ Hz, and $10^{-11}$ Hz. As $\lambda$ decreases, the CSL effect becomes weaker and weaker. In order to amplify this effect, we employ bigger and bigger spheres, $R$ from $0.15 r_c$ to $0.22 r_c$, so that $\lambda_{\rm sph}$ is enlarged while $D_t$ is not increased too much retaining a nonzero entanglement. In this process, $D_a$ is no longer negligible and starts to play its role making the solid lines in Fig.~\ref{fig3} (a)-(c) less and less linear, especially when $\omega_{1,2}$ is small. Since it has the same scaling as $\lambda_{\rm sph}$ versus $\omega_{1,2}$, i.e., $D_a$, $\lambda_{\rm sph} \propto 1/\omega_{1,2}$ (see Table~\ref{table}), its value, depending on the gas temperature and pressure, will be the most relevant factor which eventually determines the value of the upper bound of $\lambda$ that can be tested based on the system of levitated nanospheres, either optically or magnetically~\cite{OriolPRL} trapped. Another efficient way to increase $\lambda_{\rm sph}$ and meantime decrease $D_t$ is by lowering the trapping frequency $\omega_{1,2}$. However, the condition $\kappa_{\rm eff} \ll \omega_{1,2}, |\omega_1-\omega_2|$ limits the smallest value that $\omega_{1,2}$ can take, which however can be relaxed by reducing $\kappa_{\rm eff}$ implemented by taking larger values of $r_B$. Even so, one should not consider taking too small values of $\omega_{1,2}$ because when the mechanical frequencies are too low other unwanted electronic noises will enter into the system making our scheme less effective. We see in Fig.~\ref{fig3} (d)-(f) that, as $\lambda$ decreases, the distinguishability of the two curves with and without the CSL reduces. At $\lambda\sim10^{-11}$ Hz, the two curves are no longer that distinguishable (not like in (d), (e) different signs of the slope of the curves for small $\omega_{1,2}$), nevertheless, the difference of their values is still considerable (thanks to the powerful scheme~\cite{JieCF,JieNJP}). If the difference caused by all uncontrolled noises and system errors, e.g., due to the imprecision of measurements and of the calibration of the experimental parameters, is smaller than that induced by the CSL effect (the relative difference between the two curves in Fig.~\ref{fig3} (c), when $\omega_1$ is small, is about $19\%$), one can determine the CSL effect with the collapse rate down to $\lambda\sim10^{-11}$ Hz, and even lower.

We finally discuss how to detect the generated entanglement of the two nanospheres. We follow the detection scheme provided in Ref.~\cite{JieNJP}, i.e., sending into the cavity two weak red-detuned probe fields with detunings respectively equal to the two mechanical frequencies, i.e., $\Delta_j^p=\omega_{cj}-\omega_j^p=\omega_j$ ($j=1,2$ and $\omega_{cj}$ are another two cavity resonance frequencies). The probe modes adiabatically follow the dynamics of the two MRs and the output of the readout cavity $a_j^{\rm out}$ is given by~\cite{DV07}
\begin{equation}
\label{output} a_j^{\rm out}= \ii \frac{G^p_j }{\sqrt{\kappa_{\rm eff}}} b_j +
a_{j}^{\rm in},\;\;\;j=1,2,
\end{equation}
where $b_j=(x_j+\ii p_j)/\sqrt{2}$, $G^p_j$ is the very small optomechanical coupling with the probe mode and $a_{j}^{\rm in}$ denotes the input vacuum noise. Therefore, by homodyning the probe mode outputs from the transmission of the controllable beamsplitter (see Fig.~\ref{fig1}), and by changing the phases of the corresponding local oscillator, the quadratures of the two MRs, $\{x_1,p_1,x_2,p_2\}$, are measured, and thus the covariance matrix of the quadratures is constructed, from which the entanglement can then be numerically obtained in the way introduced in the Appendix. The probe fields will also induce mechanical diffusion due to the scattering of photons, however, it can be neglected if the probe fields are sufficiently weak. We have checked that in all plots of Fig.~\ref{fig3}, $D_c$ is always negligible compared to other diffusion rates, which means that even if the power of the probe field is equivalent to that of the driving laser, the diffusion induced by it is also negligible.

\section{conclusions and discussions}
\label{concl}

We have suggested a scheme based on entangled levitating nanospheres to probe the possible effect of the CSL theory. It is designed for nanospheres optically trapped in Fabry-P\'erot cavities, however, our scheme can also be applied to the system of magnetically trapped spheres, where the diffusion due to the scattering of trapping field will be significantly reduced~\cite{OriolPRL}. We have shown that the steady-state entanglement of the center-of-mass motion of the two nanospheres is particularly sensitive to the noises in the system and the CSL effect results in remarkably large reduction of the entanglement. The entanglement shows distinguishable scalings, with and without the CSL effect, with the system parameter of the trapping frequency for the collapse rate down to $\lambda \sim 10^{-10}$ Hz, while it starts to show similar scalings for $\lambda \sim 10^{-11}$ Hz, implying that our scheme can {\it unambiguously} determine the CSL effect for $\lambda$ down to $\sim 10^{-10}$ Hz with realistic parameters and can also test it for $\lambda \sim 10^{-11}$ Hz, and even lower, if the difference of the entanglement caused by uncontrolled noises and the imprecision of measurements and parameter calibration is smaller than that induced by the CSL effect. 

We notice that under the same conditions of the temperature and pressure,  $T=10$ mK and $P_a=10^{-12}$ Torr, the present scheme is not as efficient as the one provided in Ref.~\cite{Jie} where the observable is instead the phase of the cavity output light. This is because, as mentioned at the beginning in Sec.~\ref{results}, in order to have sizable entanglement a relatively small size of the sphere is adopted yielding a larger $D_a$ (since $D_a \propto 1/R$), which eventually determines this scheme, based on entanglement, could not {\it unambiguously} probe the CSL effect for $\lambda$ down to $10^{-12}$ Hz, which however can be done by the scheme of Ref.~\cite{Jie}. One promising solution is to employ the system of magnetically trapped spheres, e.g., the proposal provided in Ref.~\cite{OriolPRL}, where a superconducting microsphere is magnetically trapped close to a quantum circuit. In contrast with optical levitation, the main diffusion due to the scattering of trapping light is absent and the decoherence in the magnetic levitation system is predicted to be very small. The new system brings in additional sources of noise, e.g., due to hysteresis losses in the superconducting coils, fluctuations in the trap frequency and trap center, however, all of them are predicted to be negligible~\cite{OriolPRL}. Adopting such a magnetic levitation scheme, the entanglement could be generated with (much) larger size of the spheres, which will significantly increase $\lambda_{\rm sph}$ and meantime reduce $D_a$. The CSL effect with $\lambda$ well below $10^{-12}$ Hz, if is present, is expected to be probed with realistic parameters.

\section*{ACKNOWLEDGMENT}

This work has been supported by the Major State Basic Research Development Program of China (Grant No. 2012CB921601), the National Natural Science Foundation of China (Grant Nos: 11504218, 11634008, 11674203, 91336107, 61227902).

\section*{APPENDIX}

Here we provide the details how we obtain the steady-state entanglement between the two MRs. The entanglement is calculated based on the covariance matrix of the two mechanical modes. The covariance matrix can be achieved by solving the quantum Langevin equations \eqref{QLEs}, which can be rewritten in the following form
\begin{equation}
\dot{U} (t) = {\cal A} \, U(t) + {\cal N}(t),
\label{AppenEq}
\end{equation}
where $U(t)$ is the vector of quadrature fluctuation operators of the two mechanical modes and one cavity mode, i.e., $U(t)=\big (x_1(t),  p_1(t),  x_2(t),  p_2(t),  X(t),  Y(t) \big )^{\rm T}$. ${\cal A}$ is the drift matrix, which takes the form of 
\begin{equation}
{\cal A}=
\begin{pmatrix}
-\frac{\gamma}{2} & 0  & 0  & 0 & 0  & -G_1 \\
0 & -\frac{\gamma}{2}  &  0  & 0  & -G_1 & 0  \\
0 &  0  &  -\frac{\gamma}{2}  & 0 &  0  &  G_2 \\
0 &  0  &  0  & -\frac{\gamma}{2}  &  -G_2  & 0  \\  
0 &  -G_1  & 0  & G_2  & -\kappa_{\rm eff}  & \Delta \\
-G_1 &  0  & -G_2  & 0 & -\Delta  & -\kappa_{\rm eff}  \\
\end{pmatrix},
\label{drift}
\end{equation}
where $\Delta$ is the effective detuning (its exact expression is provided in Ref.~\cite{JieCF})  and we take $\Delta=0$ corresponding to the optimal detuning for the entanglement. The system is stable when all the eigenvalues of the drift matrix ${\cal A}$ have negative real parts, which can be simply achieved when $|G_1|<|G_2|$ is fulfilled~\cite{JieCF}. ${\cal N}(t)$ is the vector of noise quadrature operators associated with the noise terms in the equations \eqref{QLEs}. 

The {\it steady-state} covariance matrix $V(t {\to} \infty)$ of the system quadratures, with its entries defined as $V_{ij}=\frac{1}{2}\av{\{ U_i,U_j \} }$  ($\{\cdot,\cdot\}$ denotes an anticommutator, and $i,j=1,2,...,6$), is obtained by solving the Lyapunov equation
\begin{equation}
{\cal A} V + V {\cal A}^{\rm T}= -{\cal D},
\end{equation}
where ${\cal D}$ is the diffusion matrix, with its entries defined as 
\begin{equation}
\frac{1}{2}\av{{\cal N}_i(t) {\cal N}_j(s)+{\cal N}_j(s){\cal N}_i(t)}={\cal D}_{ij} \delta(t-s).
\end{equation}
The diffusion matrix is a diagonal matrix, which is ${\cal D}={\rm diag} \big[ \, (D_t^1 {+} D_c^1 {+} D_a^1 {+}\lambda_{\rm sph}^1)/2, \, (D_t^1 {+} D_c^1 {+} D_a^1 {+}\lambda_{\rm sph}^1)/2, \, (D_t^2 {+} D_c^2 + D_a^2 {+} \lambda_{\rm sph}^2)/2,  (D_t^2 {+} D_c^2 {+} D_a^2 {+}\lambda_{\rm sph}^2)/2, \kappa_{\rm eff}, \kappa_{\rm eff}  \big]$. 

Once the covariance matrix $V$ is obtained, the entanglement can then be quantified by means of the logarithmic negativity~\cite{Jens}: 
\begin{equation}
E_N=\max[0,-\ln2\tilde\nu_-], 
\end{equation}
where $\tilde\nu_-=\min{\rm eig}|i\Omega_2\tilde{V}_m|$  ($\Omega_2{=}\oplus^2_{j=1}i\sigma_y$ the so-called symplectic matrix and $\sigma_y$ the $y$-Pauli matrix) is the minimum symplectic eigenvalue of the covariance matrix $\tilde{V}_m={\cal P}{V_m}{\cal P}$, with $V_m$ the $4\times 4$ covariance matrix associated with the two mechanical modes and ${\cal P}={\rm diag}(1,1,1,-1)$ the matrix that inverts the sign of momentum of the 2nd MR, i.e., $p_2 \to - p_2$, realizing partial transposition at the level of covariance matrices~\cite{Simon}.


\begin{thebibliography}{99}



\bibitem{BassiRMP}A. Bassi, K. Lochan, S. Satin, T. P. Singh, and H. Ulbricht, Rev. Mod. Phys. {\bf 85}, 471 (2013).

\bibitem{Bassi}A. Bassi and G. C. Ghirardi, Physics Reports {\bf 379}, 257 (2003); S. L. Adler and A. Bassi, Science {\bf 325}, 275 (2009).

\bibitem{GRW}G. C. Ghirardi, A. Rimini, and T. Weber, Phys. Rev. D {\bf 34}, 470 (1986).

\bibitem{CSL}G. C. Ghirardi, P. Pearle, and A. Rimini, Phys. Rev. A {\bf 42}, 78 (1990); G. C. Ghirardi, R. Grassi, and F. Benatti, Found. Phys. {\bf 25}, 5 (1995).

\bibitem{DP}L. Di\'osi, Phys. Lett. A {\bf 120}, 377 (1987); L. Di\'osi, Phys. Rev. A {\bf 40}, 1165 (1989); R. Penrose, Gen. Rel. Grav. {\bf 28}, 581 (1996).



\bibitem{ArndtRMP}K. Hornberger, S. Gerlich, P. Haslinger, S. Nimmrichter, and M. Arndt, Rev. Mod. Phys. {\bf 84}, 157 (2012).

\bibitem{cluster} P. Haslinger, N. D\"orre, P. Geyer, J. Rodewald, S. Nimmrichter, and M. Arndt, Nat. Phys. {\bf 9}, 144 (2013).



\bibitem{Mauro}M. Bahrami, M. Paternostro, A. Bassi, and H. Ulbricht, Phys. Rev. Lett. {\bf 112}, 210404 (2014).

\bibitem{Nimmrichter}S. Nimmrichter, K. Hornberger, and K. Hammerer, Phys. Rev. Lett. {\bf 113}, 020405 (2014).

\bibitem{Sekatski}P. Sekatski, M. Aspelmeyer, and N. Sangouard, Phys. Rev. Lett. {\bf 112}, 080502 (2014); M. Ho, A. Lafont, N. Sangouard, and P. Sekatski, New J. Phys. \textbf{18}, 033025 (2016).

\bibitem{Diosi}L. Di\'osi, Phys. Rev. Lett. {\bf 114}, 050403 (2015).

\bibitem{Jie}J. Li, S. Zippilli, J. Zhang, and D. Vitali,  Phys. Rev. A {\bf 93}, 050102(R) (2016).

\bibitem{Barker0}D. Goldwater, M. Paternostro, and P. F. Barker, Phys. Rev. A {\bf 94}, 010104(R) (2016).


\bibitem{OMRMP}M. Aspelmeyer, T. J. Kippenberg, and F. Marquardt, Rev. Mod. Phys. {\bf 86}, 1391 (2014).


\bibitem{mirror} W. Marshall, C. Simon, R. Penrose, and D. Bouwmeester, Phys. Rev. Lett. {\bf 91}, 130401 (2003); B. Pepper, R. Ghobadi, E. Jeffrey, C. Simon, and D. Bouwmeester, Phys. Rev. Lett. {\bf 109}, 023601 (2012).

\bibitem{nanosphere}O. Romero-Isart, A. C. Pflanzer, F. Blaser, R. Kaltenbaek, N. Kiesel, M. Aspelmeyer, and J. I. Cirac, Phys. Rev. Lett. {\bf 107}, 020405 (2011); O. Romero-Isart, Phys. Rev. A {\bf 84}, 052121 (2011).

\bibitem{Ulbricht}J. Bateman, S. Nimmrichter, K. Hornberger, and H. Ulbricht, Nature Comm. {\bf 5}, 4788 (2014).

\bibitem{Wan}C. Wan,  M. Scala, G. W. Morley, A. A. Rahman, H. Ulbricht, J. Bateman, P. F. Baker, S. Bose, and M. S. Kim, Phys. Rev. Lett. {\bf 117}, 143003 (2016).


\bibitem{Serafini}M. G. Genoni, O. S. Duarte, and A. Serafini,  New J. Phys. {\bf 18}, 103040 (2016).

\bibitem{Millen}S. McMillen, M. Brunelli, M. Carlesso, A. Bassi, H. Ulbricht, M. G. A. Paris, and M. Paternostro, arXiv:1606.00070.


\bibitem{PRL02}
S. Mancini, V. Giovannetti, D. Vitali and P. Tombesi, Phys. Rev. Lett. \textbf{88}, 120401 (2002).

\bibitem{jopa}D. Vitali, S. Mancini, and P. Tombesi, J. Phys. A: Math. Theor. \textbf{40}, 8055 (2007).

\bibitem{genesNJP}C. Genes, D. Vitali, and P. Tombesi, New J. Phys. \textbf{10}, 095009
(2008).

\bibitem{hartmann}M. J. Hartmann and M. B. Plenio, Phys. Rev. Lett. \textbf{101}, 200503
(2008).

\bibitem{Peng03}
J. Zhang, K. C. Peng, and S. L. Braunstein, Phys. Rev. A \textbf{68}, 013808(2003).

\bibitem{Jie13}
J. Li, S. Gr\"oblacher and M. Paternostro, New J. Phys. {\bf 15}, 033023 (2013).

\bibitem{Ge13}
W. Ge, M. Al-Amri, H. Nha, and M. S. Zubairy, Phys. Rev. A {\bf 88}, 022338 (2013).


\bibitem{EPL}
M. Pinard, A. Dantan, D. Vitali, O. Arcizet, T. Briant and A. Heidmann, Europhys. Lett. \textbf{72}, 747 (2005).

\bibitem{entswap}
S. Pirandola, D. Vitali, P. Tombesi, S. Lloyd, Phys. Rev. Lett. \textbf{97}, 150403 (2006).

\bibitem{bjorke}
K. Borkje, A. Nunnenkamp, and S. M. Girvin, Phys. Rev. Lett. \textbf{107}, 123601 (2011).

\bibitem{mehdi1}
M. Abdi, S. Pirandola, P. Tombesi, and D. Vitali, Phys. Rev. Lett. \textbf{109}, 143601 (2012).

\bibitem{woolley}
M. J. Woolley and A. A. Clerk, Phys. Rev. A \textbf{87}, 063846 (2013).

\bibitem{mehdi2}
M. Abdi, S. Pirandola, P. Tombesi, and D. Vitali, Phys. Rev. A \textbf{89}, 022331 (2014).

\bibitem{Savona}
H. Flayac and V. Savona, Phys. Rev. Lett. {\bf 113}, 143603 (2014).


\bibitem{Clerk}
Y.-D. Wang and A. A. Clerk, Phys. Rev. Lett. \textbf{108}, 153603 (2012).

\bibitem{Tan}
H. Tan, G. Li, and P. Meystre, Phys. Rev. A \textbf{87}, 033829 (2013).

\bibitem{WoolleyClerk}
M. J. Woolley and A. A. Clerk, Phys. Rev. A \textbf{89}, 063805 (2014).

\bibitem{Abdi2}
M. Abdi, and M. J. Hartmann, New J. Phys. \textbf{17}, 013056 (2015).

\bibitem{Buchmann}
L. F. Buchmann and D. M. Stamper-Kurn, Phys. Rev. A {\bf 92}, 013851 (2015).

\bibitem{JieNJP}J. Li, I. Moaddel Haghighi, N. Malossi, S. Zippilli, and D. Vitali, New J. Phys. {\bf 17}, 103037 (2015).

\bibitem{JieCF}J. Li, G. Li, S. Zippilli, D. Vitali, and T. C. Zhang, arXiv:1610.07261.



\bibitem{Belli}S. Belli, R. Bonsignori, G. D'Auria, L. Fant, M. Martini, S. Peirone, S. Donadi, and A. Bassi, Phys. Rev. A {\bf 94}, 012108 (2016).

\bibitem{Singh}S. Banerjee, S. Bera, and T. P. Singh, Phys. Lett. A {\bf 380}, 3778 (2016).


\bibitem{Chang} D. E. Chang, C. A. Regal, S. B. Papp, D. J. Wilson, J. Ye, O. Painter, H.J. Kimble, and P. Zoller, Proc. Natl. Acad. Sci. USA {\bf 107}, 1005 (2010).

\bibitem{Barker}P. F. Barker and M. N. Shneider, Phys. Rev. A {\bf 81}, 023826 (2010); J. Millen, P. Z. G. Fonseca, T. Mavrogordatos, T. S. Monteiro, and P. F. Barker, Phys. Rev. Lett. {\bf 114}, 123602 (2015).

\bibitem{Raizen}T. Li, S. Kheifets, and M. G. Raizen, Nature Phys. {\bf 7}, 527 (2011).

\bibitem{gieseler}J. Gieseler, B. Deutsch, R. Quidant, and L. Novotny, Phys.
Rev. Lett. \textbf{109}, 103603 (2012).

\bibitem{Markus}N. Kiesel, F. Blaser, U. Delic, D. Grass, R. Kaltenbaek, and M. Aspelmeyer, Proc. Natl. Acad. Sci. USA {\bf 110}, 14180 (2013).



\bibitem{Oriol2011} O. Romero-Isart, A. C. Pflanzer, M. L. Juan, R. Quidant, N. Kiesel, M. Aspelmeyer, and J. I. Cirac, Phys. Rev. A {\bf 83}, 013803 (2011).

\bibitem{Pflanzer}A. C. Pflanzer, O. Romero-Isart, and J. Ignacio Cirac, Phys. Rev. A {\bf 86}, 013802 (2012).


\bibitem{Adler}S. L. Adler, J. Phys. A: Math. Theor. {\bf 40}, 2935 (2007).

\bibitem{Vinante}A. Vinante, M. Bahrami, A. Bassi, O. Usenko, G. Wijts, and T. H. Oosterkamp, Phys. Rev. Lett. \textbf{116}, 090402 (2016).

\bibitem{coldatoms} M. Bilardello, S. Donadi, A. Vinante, and A. Bassi, Physica A {\bf 462}, 764 (2016).

\bibitem{Chen}B. Helou, B. Slagmolen, D. E. McClelland, and Y. Chen, arXiv:1606.03637.

\bibitem{Vinante2}M. Carlesso, A. Bassi, P. Falferi, and A. Vinante, arXiv:1606.04581.

\bibitem{VinanteNEW}A. Vinante, R. Mezzena, and P. Falferi, arXiv:1611.09776.

\bibitem{Pearle2014}See footnote 7 in F. Lalo\"e, W. J. Mullin, and P. Pearle, Phys. Rev. A {\bf 90}, 052119 (2014).

\bibitem{Curceanu}C. Curceanu, B. C. Hiesmayr, and K. Piscicchia, J. Adv. Phys. {\bf 4}, 263 (2015).

\bibitem{OriolPRL}O. Romero-Isart, L. Clemente, C. Navau, A. Sanchez, and J. I. Cirac, Phys. Rev. Lett. {\bf 109}, 147205 (2012).

\bibitem{DV07}D. Vitali {\it et al.}, Phys. Rev. Lett. {\bf 98}, 030405 (2007).



\bibitem{Jens}
J. Eisert, PhD Thesis, University of Potsdam, 2001; G. Vidal and R. F. Werner, Phys. Rev. A {\bf 65}, 032314 (2002); M. B. Plenio, Phys. Rev. Lett. {\bf 95}, 090503 (2005).

\bibitem{Simon}
R. Simon, Phys. Rev. Lett. {\bf 84}, 2726 (2000).



\end{thebibliography}
\end{document}